\documentclass[showpacs,preprintnumbers,amsmath,amssymb]{revtex4}
\usepackage{graphicx}% Include figure files
\usepackage{dcolumn}% Align table columns on decimal point
\usepackage{bm}% bold math

\begin{document}
\title{ Unified Solution of the Expected Maximum of a Discrete Time Random Walk and
the Discrete Flux to a Spherical Trap}

\author {Satya N. Majumdar $^{1}$, Alain Comtet $^{1,2}$ and Robert M. Ziff $^{3,}$*}
\address{
{\small $^1$ Laboratoire de Physique Th\'eorique et Mod\`eles Statistiques,
        Universit\'e Paris-Sud. B\^at. 100. 91405 Orsay Cedex. France}\\
{\small $^2$ Institut Henri Poincar\'e, 11 rue Pierre et Marie Curie, 75005 Paris, France}\\
{\small $^3$ Michigan Center for
Theoretical Physics and Department of Chemical
Engineering, University of Michigan, Ann Arbor, MI USA 48109-2136} 
}

%\maketitle

\date{\today}

\begin{abstract}
Two random-walk related problems which have been studied independently
in the past, the expected maximum of a random walker in one dimension and the flux to a
spherical trap of particles undergoing discrete jumps in three dimensions, are shown to be
closely related to each other and
are studied using a unified approach as a solution to a Wiener-Hopf
problem.  For the flux problem, this work shows that a constant
$c = 0.29795219$ which appeared in the context of the 
boundary extrapolation length, and was
previously found only
numerically, can be derived analytically.  The same constant enters in
higher-order corrections to the expected-maximum asymptotics.
As a byproduct, we also prove a new universal result in the context of the
flux problem which is an analogue of the Sparre Andersen theorem proved in
the context of the random walker's maximum.
\bigskip

\noindent* Corresponding author, rziff$@$umich.edu, 734-764-5498, 734-763-0459 (fax) 

\noindent {\bf Running title:} Maximum of Random Walk and Flux to a Trap

\end{abstract}
\maketitle

\section{Introduction}

Random walks arise in an astounding variety of problems in physics
as well as mathematics, computer science, etc., and
great progress has been made in solving many of their deep and subtle
properties.
Two seemingly unrelated problems, the expected maximum of a random
walker in {\em one} dimension undergoing jumps drawn from a {\em uniform} 
distribution, and the flux to a trap of particles undergoing random walks
in {\em three} dimensions, have
been studied independently over the last several years, and their 
solutions seem to involve a similar numerical constant 0.29795219,
a coincidence that has not been noticed before.
The similarity of these constants
suggests that these two problems may be intimately related.
For the first problem, the constant was first computed numerically
by evaluating a rather complicated double series expansion~\cite{CS,CFFH}
and very recently, an exact closed form expression of the constant
has been found~\cite{CM} that is valid not just for the uniform 
jump distribution, but for any arbitrary continuous and symmetric jump distribution.
For the second problem of flux to a spherical trap, the 
corresponding constant was computed only numerically~\cite{Ziff1}. Therefore, finding
the relation between the two problems raises the possibility that the
flux problem can be solved analytically for the first time.  Indeed,
this is what we accomplish in this paper.

The two problems we consider are:
\vspace{0.5cm}
   
{\bf Problem I: The asymptotic behavior
of the expected maximum position of a discrete time random walker moving on a continuous line.} The
position $x_n$ of the walker after $n$ steps evolves for $n\ge 1$ via,
\begin{equation}
x_n = x_{n-1}+ \xi_n
\label{evol1}
\end{equation}
starting at $x_0=0$, where the step lengths $\xi_n$'s are independent and identically distributed
(i.i.d) random variables with zero mean and each drawn from the same probability distribution,
${\rm Prob}(\xi_n\le x)=\int_{-\infty}^x f(y)\,dy$,  $f(x)$ being a continuous and
symmetric probability density normalized to unity.
Let $M_n$ denote the positive maximum of the random walk up to $n$ steps (see Fig. 1),
\begin{equation}
M_n = {\rm max} (0, x_1, x_2, \dots, x_n).
\label{max1}
\end{equation}

We are interested in the asymptotic large-$n$ behavior of the expected maximum $E(M_n)$.
This question arose some years ago in the context of a packing problem in two dimensions
where $n$ rectangles of variable sizes
are packed in a semi-infinite strip of width one~\cite{CS,CFFH}.
It was shown in Ref.\ \cite{CFFH} that for the special case of the uniform jump distribution,
$f(x)=1/2$ for $-1\le x \le 1$ and $f(x)=0$ outside, for large $n$,
\begin{equation}
E[M_n] = \sqrt{\frac{2n}{3\pi}} -0.29795219028\dots + O(n^{-1/2}).
\label{flaj1}
\end{equation}
The leading $\sqrt{n}$ behavior is easy to understand and can be derived from the corresponding
behavior of a continuous-time Brownian motion after a suitable rescaling~\cite{CFFH}. However,
the leading finite-size correction term turns out to be a nontrivial constant $-c$ with
$c=0.29795219028\dots$ that
was computed in Ref.\ \cite{CFFH} by enumerating a somewhat intricate double series obtained after
a lengthy calculation. Recently, it was shown~\cite{CM} that for arbitrary continuous and symmetric 
jump 
distribution
$f(x)$ with a finite second moment $\sigma^2= \int_0^{\infty} x^2 f(x) dx$, the expected
maximum has a similar asymptotic behavior as in the uniform case, namely,
\begin{equation}
E[M_n]= \sigma \sqrt{\frac{2n}{\pi}} - c + O(n^{-1/2}).
\label{cm1}
\end{equation}
Moreover, an exact expression for the constant $c$ was found~\cite{CM}
\begin{equation}
c = - \frac{1}{\pi}\, \int_0^{\infty} \frac{dk}{k^2}\, \ln \left[\frac{1-{\hat 
f}(k)}{\sigma^2k^2/2}\right],
\label{c1}
\end{equation}
where ${\hat f}(k) = \int_{-\infty}^{\infty} f(x)\, e^{ikx}\, dx$ is the Fourier transform
of $f(x)$. In particular, for the uniform distribution, $f(x)=1/2$ for $-1\le x \le 1$ and $f(x)=0$ 
outside, one has ${\hat f}(k)= \sin (k)/k$ and one gets from Eq.\  (\ref{c1}) an exact expression,
\begin{equation}
c = - \frac{1}{\pi}\, \int_0^{\infty} \frac{dk}{k^2}\, \ln \left[\frac{6}{k^2}\left(1-\frac{\sin 
k}{k}\right)\right] = 0.29795219028\dots   
\label{c2}
\end{equation}

\begin{figure} \includegraphics[width=4in]{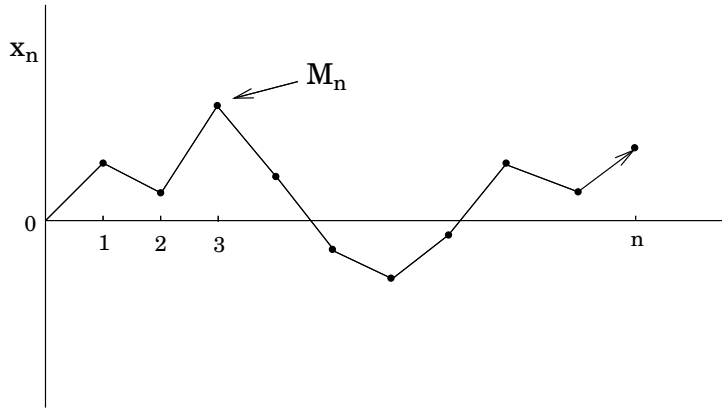} \caption{\label{fig:maxfig}A typical
configuration of a random walker in one dimension up to $n$ steps, starting at $0$ at $n=0$ with $M_n$
denoting the maximum up to $n$ steps.} \end{figure}

%\begin{figure}[htbp] 
%\epsfxsize=8cm 
%\centerline{\epsfbox{maxfig.eps}} 
%\caption{A typical configuration of a random walker in one dimension upto $n$ steps, 
%starting at $0$ at $n=0$ with
%$M_n$ denoting the maximum up to $n$ steps.} %\label{fig:maxfig} %\end{figure}

\vspace{0.5cm}

{\bf Problem II: The calculation of flux to a spherical trap
in three dimensions.} Consider first the classic Smoluchowski problem~\cite{Smolu} where point 
particles
are initially distributed uniformly with density $\rho_0$ outside a sphere of radius $R$ in three dimensions.
Each particle subsequently performs continuous-time Brownian motion with a diffusion
constant $D$, independent of each other.
One is interested in computing the flux of particles $\Phi(t)$ to the sphere at time $t$.
This can be done by solving the diffusion equation outside the sphere with an absorbing boundary 
condition on the surface of the sphere and the result is well 
known~\cite{Smolu,Chandra,Weiss}. One gets
\begin{equation}
\Phi(t) = 4\pi R D \rho_0 \left[1+ \frac{R}{\sqrt{\pi\, D\, t}} \right].
\label{phi1}
\end{equation}
valid for all $t > 0$. Also, as $t\to \infty$, the density profile
outside the sphere becomes time independent and has a simple form
\begin{equation}
\rho(r) =\rho_0 \left(1-\frac{R}{r}\right)
\label{denprof1}
\end{equation}
for all $r\ge R$. Far from the sphere the density remains unchanged from its initial
value $\rho_0$ and as one approaches the surface of the sphere, the density vanishes. 

An interesting issue, first studied in Ref.\ \cite{Ziff1}, is how do the 
steady-state profile in Eq. (\ref{denprof1}) and correspondingly the expression of flux
in Eq.\  (\ref{phi1}) get modified when each of the point particles, instead of 
performing continuous-time Brownian 
diffusion, undergoes discrete `Rayleigh flights', i.e. a particle jumps, at every discrete
time step $\tau$, a fixed step length $l$ whose direction is chosen arbitrarily in the 
three-dimensional space (see Fig. 2). 
In Ref.\ \cite{Ziff1}, it was shown that the expression for the flux, at late times $t=n\tau$
and for $0< l\le 2R$, now gets replaced by
\begin{equation}
\Phi(t) = 4\pi (R-c'\,l) D' \rho_0  \left[1+ \frac{R-c'\, l}{\sqrt{\pi\, D'\, t}} + O(t^{-3/2})\right],
\label{phi2}
\end{equation}
where $D'= l^2/6\tau$ and $c'$ is a constant whose numerical value was obtained by 
an iterative numerical solution of the density profile, with the result
\begin{equation}
c' \approx 0.29795219
\label{c'1}
\end{equation}
The density profile $\rho_n(r)$ after $n$ steps also gets modified rather drastically, as found
numerically in Ref.\ \cite{Ziff1}. In particular, the steady-state density profile
$\rho_{\infty}(r)$ in the discrete problem turns out to be quite different
from its continuous-time counterpart. While very far away 
from the sphere the steady-state density profile behaves as
\begin{equation}
\rho_{\infty}(r>>R) =\rho_0 \left(1-\frac{R-c' l}{r}\right)
\label{denprof2}
\end{equation}
where $c'$ is as in Eq. (\ref{c'1}), the density actually tends to a nonzero
constant as one approaches the surface of the sphere from outside
\begin{equation}
\rho_{\infty}(r\to R) = 0.408245\, \frac{\rho_0 l}{R}
\label{densurf1}
\end{equation}
where the constant $0.408245$ was evaluated numerically in Ref.\ \cite{Ziff1}.
This is in stark contrast to the continuous-time Brownian case where the density
vanishes on the surface of the sphere.

The distance $c' \, l$ is the `Milne extrapolation length'~\cite{BT,CZ,Williams} and 
represents the 
distance inside the surface where the far steady steady-state solution in Eq. (\ref{denprof2}) 
extrapolates to $\rho_{\infty}=0$. This steady-state density distribution implies the leading-order term of the 
flux given in Eq.\ (\ref{phi2}). Thus at late times, the continuum formula for the flux in 
Eq.\ (\ref{phi1})
still remains valid for the discrete jump case, but with an effectively 
smaller radius $R-c'\, l$ of the trap
as in Eq.\ (\ref{phi2}). Thus the effect on the flux due to the discrete nature of the jumps, 
at least at late times, is simply
to renormalize the radius of the trap to a smaller value.

\begin{figure}
\includegraphics[width=4in]{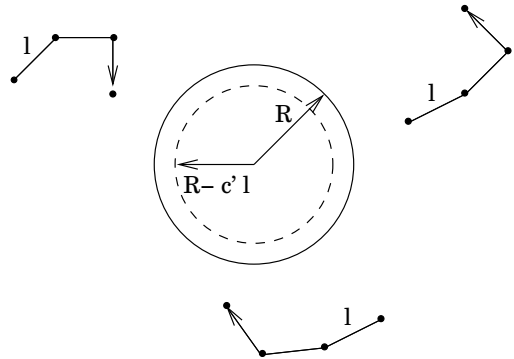}
\caption{\label{fig: fluxfig} Independent Rayleigh flights in three dimensions in presence of a 
spherical trap with %radius R. 
The discreteness of the jumps shows up 
effectively in a renormalized sphere with a smaller %radius $R-c'\,l$
where $c'\, l$ is the Milne extrapolation length}
\end{figure}

%\begin{figure}[htbp]
%\epsfxsize=8cm
%\centerline{\epsfbox{fluxfig.eps}}
%\caption{Independent Rayleigh flights in three dimensions in presence of a spherical trap with %radius R. 
%The discreteness of the jumps shows up effectively in a renormalized sphere with a smaller %radius $R-c'\,l$
%where $c'\, l$ is the Milne extrapolation length.}
%\label{fig:fluxfig}
%\end{figure}

Comparing Eqs.\ (\ref{c2}) and (\ref{c'1}) one finds, rather amazingly, that the two constants 
$c$ and $c'$, in these two {\it a priori} unrelated problems, are identical at least up to $8$ decimal places!
This raises an interesting question: are they equal? In this paper, we indeed prove that
$c=c'$. In the process, we also provide exact solutions to many other features of the
flux problem that were observed numerically in Ref.\ \cite{Ziff1}. 
For example, we will calculate exactly the steady-state density profile and will
prove that indeed it approaches to a constant on the surface of the sphere as in Eq. 
(\ref{densurf1}) and the constant $0.408245$ is actually $1/\sqrt{6}$.
Our method consists in
showing that both of these problems can be cast into the same Wiener-Hopf type
problem involving an integral equation over half-space, albeit with different initial conditions. We then 
obtain explicit solutions
to this Wiener-Hopf problem with these two different initial conditions. The general solution
turns out to be a product of two parts, one that explicitly depends on the initial condition
and the other part which is a `Green's function' that is independent of the initial condition.
The constant $c$, given by the exact expression in Eq.\ (\ref{c2}), is part of this
`initial condition independent' Green's function and hence it appears in both problems.

\section{Maximum of a Random Walker as a Wiener-Hopf problem}

We consider a discrete time random walker hopping on a continuous line. The position $x_n$
of the walker evolves via Eq.\ (\ref{evol1}), starting at $x_0=0$. The maximum $M_n$, defined
in Eq.\ (\ref{max1}), is a random variable. Let $q_n(z)={\rm Prob}(M_n \le z)$ denote
the cumulative distribution of the maximum and $q_n'(z)=dq_n/dz$ its probability
density with $z\in [0,\infty]$.  Then, a simple integration by parts gives
\begin{equation}
E[M_n]= \int_0^{\infty} z q_n'(z)dz =-\int_0^{\infty}z \frac{d}{dz}(1-q_n(z))\, dz =\int_0^{\infty} 
(1-q_n(z))\, dz,
\label{emn1}
\end{equation}
where we have used the normalization condition, $q_n(\infty)=1$.
Using translational invariance and the fact that the jump distribution is 
symmetric, one can also interpret 
$q_n(z)$ as the probability that
a random walker, starting initially at position $z>0$, stays positive up to step $n$. Then it is
easy to write down, using the Markov property of the evolution in Eq.\ (\ref{evol1}), the
following recurrence relation~\cite{CM}, valid for all $z\ge 0$,
\begin{equation}
q_n(z)= \int_0^{\infty} q_{n-1}(z') f(z-z')\, dz', \quad\quad {\rm starting\,\, with}\,\,\, q_0(z)=1.
\label{hop1}
\end{equation}
The generating function ${\tilde q}(z,s)= \sum_{n=0}^{\infty} q_n(z) s^n $ then satisfies
an inhomogeneous Wiener-Hopf integral equation~\cite{CM} 
\begin{equation}
{\tilde q}(z,s) = s\, \int_0^{\infty} {\tilde q} (z',s) f(z-z') dz' + q_0(z),
\label{hop2}
\end{equation}
where the inhomogeneous term $q_0(z)=1$ arises from the initial condition.
We need to thus solve this integral equation to obtain the full probability
distribution $q_n(z)$ of the maximum. The mean value $E[M_n]$ can 
then be computed from Eq.\ (\ref{emn1}).

\section{Flux to a trap as a Wiener-Hopf problem}

We consider a sphere of radius $R$ in three dimensions. Outside the sphere point particles are
initially distributed with uniform density $\rho_0$. Particles subsequently perform
independent Rayleigh flights, i.e. at every time step $\tau$, each particle, independently
of others, jumps a fixed distance $l$ in a direction chosen randomly. The object of
interest is the flux at late times $t$ to the sphere. For simplicity, we assume
$\tau=1$ (so that the jumps occur at integer steps) and also $l=1$. 
Since the particles are independent, one can alternately think of a single 
particle whose probability density $\rho_n(\vec r)$ at $\vec r$ after $n$ steps evolves
via the Markov equation~\cite{Ziff1}
\begin{equation}
\rho_n(\vec r) = \int W(\vec r|\,{\vec r}\,') \rho_{n-1}({\vec r}\,') d {\vec r}\,'
\label{fop1}
\end{equation}
where $W(\vec r|\, {\vec r}\,')= \delta(|\vec r -{\vec r}\,'|-1)/{4\pi}$ is the jump
probability density at each step from $\vec r\,'$ to $\vec r$ and the integral in Eq.\ (\ref{fop1})
extends over the full three-dimensional space outside the sphere of radius $R$. The recursion 
relation in Eq.\ (\ref{fop1})
starts with the initial condition, $\rho_0(\vec r)= \rho_0$ for $r>R$ and $\rho_0(\vec r)=0$
for $r\le R$. Since the initial condition is spherically symmetric, it is clear that
from Eq.\ (\ref{fop1}) that this symmetry will be maintained at all $n$, implying
$\rho_n(\vec r) =\rho_n(r)$. Thanks to this spherical symmetry, the three-dimensional
problem thus becomes a one-dimensional problem where one considers only the
radial direction after integrating out the angular coordinates. Defining $P_n(r)\equiv 
4\pi r^2 \rho_n(r)$ as the probability
that the particle is in the shell $[r, r+dr]$ after $n$ steps, it follows that
$P_n(r)$ evolves via the recurrence equation
\begin{equation}
P_n(r) = \int w(r|\,r') P_{n-1}(r') dr' \quad\quad {\rm starting\,\, with}\,\,\, P_0(r)=4\pi r^2 \rho_0 
\theta(r-R)
\label{fop2}
\end{equation}
where $\theta(x)$ is the Heaviside theta function and
$w(r|\,r')$ is the jump probability density from radius $r'$ to $r$, which can be easily
calculated by integrating the kernel $W(\vec r|\, \vec r')$ over the angular 
coordinates~\cite{Ziff1,CK}
\begin{eqnarray}
w(r|\,r')dr &=& \frac{r\, dr}{2r'}\quad\quad {\rm for} \,\,\, |r'-1|< r< r'+1 \nonumber \\
          &=& 0\quad\quad {\rm otherwise}.
\label{kernel1}
\end{eqnarray}  
To simplify, one introduces a new quantity $F_n(r)= P_n(r)/{4\pi\rho_0 r}$. The recursion
relation for $F_n(r)$, upon substituting 
the explicit form of $w(r|r')$ from Eq.\ (\ref{kernel1}) in Eq.\ (\ref{fop2}), then simplifies
\begin{eqnarray}
F_n(r)& =& \frac{1}{2}\, \int_{{\rm max}(R, r-1)}^{r+1} F_{n-1}(r') dr' \nonumber \\
      & = & \int_R^{\infty} F_{n-1}(r') f(r-r') dr' 
\label{fop3}
\end{eqnarray}
where $f(y)$ corresponds to the uniform probability density over the interval $y\in [-1,1]$, i.e.
$f(y)=1/2$ if $-1\le y\le 1$ and $f(y)=0$ otherwise. The recursion in Eq.\ (\ref{fop3})
now starts with the initial condition, $F_0(r)= r \theta(r-R)$. One can simplify Eq.\ 
(\ref{fop3}) further
by introducing a shift, i.e. defining $z=r-R$ and writing $F_n(r)= F_n(z+R)=Q_n(z)$, Eq.\ (\ref{fop3})
becomes, for all $z>0$,
\begin{equation}
Q_n(z)= \int_0^{\infty} Q_{n-1}(z') f(z-z')\, dz', \quad\quad {\rm starting\,\, with}\,\,\, Q_0(z)=R+z
\label{fop4}
\end{equation}
Defining the generating function, ${\tilde Q}(z,s)= \sum_{n=0}^{\infty} Q_n(z) s^n$, one obtains
an identical Wiener-Hopf integral equation as in Eq.\ (\ref{hop2}),
\begin{equation}
{\tilde Q}(z,s) =s\, \int_0^{\infty} {\tilde Q}(z',s) f(z-z') dz' + Q_0(z),
\label{fop5}
\end{equation}
the only difference is in the inhomogeneous term $Q_0(z)=R+z$ that is set by the initial
condition. One then needs to solve this integral equation to obtain $Q_n(z)$, from which
one can read off the density profile at step $n$
\begin{equation}
\rho_n(r) =\frac{P_n(r)}{4\pi r^2}=\frac{\rho_0}{r} F_n(r)= \frac{\rho_0}{r} Q_n(r-R).
\label{denprof3}
\end{equation}
The flux to the sphere at time step $n$ can then be computed
from the following relation~\cite{Ziff1}
\begin{eqnarray}
\Phi(n)&=& \int_{R}^{\infty} dr' \int_0^{R} dr w(r|\, r') P_{n-1}(r') \label{f1} \\
       &=& \pi \rho_0 \, \int_R^{R+1} [R^2-(r'-1)^2] F_{n-1}(r') dr' \label{f2} \\
       &=& \pi \rho_0 \, \int_0^{1} [2R(1-z)-(1-z)^2]\, Q_{n-1}(z)\, dz
\label{f3}
\end{eqnarray}
In going from Eq.\ (\ref{f1}) to (\ref{f2}) we have used the explicit form of $w(r|r')$ in Eq.\ (\ref{kernel1}).

We end this section with one remark. For a continuous-time Brownian motion it is quite
standard~\cite{Chandra,Redner} that, using the transformation $\rho(r,t)={\rho_0}F(r,t)/{r}$, the 
$3$-d 
diffusion equation for the density field $\rho(r,t)$ can be reduced to a 
$1$-d diffusion equation; the same trick
naturally works for the $3$-d Shr\"odinger equation as well. 
Based on this fact, it is natural that a similar transformation $\rho_n(r)={\rho_0}F_n(r)/r$
would also work for the discrete-time problem. However, the fact, that the reduced $1$-d
problem satisfies exactly the same integral equation (albeit with a different initial
condition) with the same {\em uniform} kernel as the $1$-d maximum displacement problem, 
is hard to guess apriori without the explicit calculation as presented here.
  
\section{Wiener-Hopf problem}

We have seen from the previous sections that the two {\it a priori} different problems (I) maximum of a
random walker hoping on a line and (II) flux to a spherical trap in three dimensions can be both
recast as the same Wiener-Hopf integral equation problem, albeit with different inhomogeneous 
terms arising due to the difference in the initial conditions of the two problems. The general 
mathematical problem then
is to solve the following half-space inhomogeneous integral equation for $z>0$
\begin{equation}   
\psi(z,s) = s \int_0^{\infty} \psi(z', s)\, f(z-z') dz' + J(z).
\label{wh1}
\end{equation}
where the inhomogeneous source term $J(z)$ is different for the two problems
\begin{eqnarray}
J(z) & =& 1 \quad\quad\quad\quad \hbox{for Problem I}\label{rwp} \\
     & = & R+z \quad\quad  \hbox{for Problem II}. \label{flp}
\end{eqnarray}
Even though in both of these problems the kernel $f(z-z')$ corresponds to the uniform jump density, i.e.
\begin{eqnarray}
f(x) &=& \frac{1}{2} \quad\quad {\rm for} \,\, -1\le x \le 1 \nonumber \\
     &=& 0 \quad\quad {\rm otherwise},
\label{uni}
\end{eqnarray}
it is useful to study the integral equation (\ref{wh1}) with a general continuous and symmetric 
kernel
$f(z)=f(-z)$ that is normalized $\int_{-\infty}^{\infty} f(z) dz=1$ and has a finite second moment
$\sigma^2= \int_{-\infty}^{\infty} z^2 f(z) dz$.

The explicit solution $\psi(z,s)$ of Eq.\ (\ref{wh1}) with different source terms as
in Eqs.\ (\ref{rwp}) and (\ref{flp}) will then provide the solutions to the two problems.
In Problem I, $\psi(z,s)= {\tilde q}(z,s)=\sum_{n=0}^{\infty} q_n(z) s^n $ provides
the generating function for the cumulative distribution of the maximum of the random walk up to $n$ 
steps.
In Problem II, $\psi(z,s)= {\tilde Q}(z,s)= \sum_{n=0}^{\infty} Q_n(z) s^n $ gives the generating
function for the density profile ${\rho}_n(r)= \rho_0 Q_n(z=r-R)/r$ of the particles outside the 
sphere
of radius $R$ in  three dimensions.  
It turns out, as will be shown later explicitly, that the 
difference in the source term in Eqs.\ (\ref{rwp})
and (\ref{flp}) actually leads to completely different types of solutions to the integral equation
(\ref{wh1}). In Problem I, the solution $q_n(z)$ depends explicitly on $n$ even at late times, i.e.
for large $n$, and does not have an $n$-independent stationary solution. Rather it has a scaling 
solution involving both $z$ and $n$. In contrast, the solution $Q_n(z)$ in Problem II approaches
an $n$-independent stationary solution. 

\subsection{Explicit solution for exponential kernel}

Before providing the general solution for arbitrary 
continuous and symmetric kernel $f(z)$, it is instructive to derive the explicit solutions with the 
two
different source terms for a special kernel $f(z)=\exp[-|z|]/2$. This will clearly
bring out how the different source terms lead to different behavior of the same
integral equation. The exponential kernel $f(z)=\exp[-|z|]/2$ is special since one can recast the
integral equation (\ref{wh1}) into a differential equation by using the identity
$f''(z)= f(z)-\delta(z)$, where $f''(z)=d^2f/dz^2$. Differentiating Eq.\ (\ref{wh1}) twice
with respect to $z$ and using the above identity, one gets for all $z>0$
\begin{equation}
\frac{d^2\psi}{dz^2} = (1-s) \psi(z,s) + J''(z)-J(z).
\label{exp1}
\end{equation}

Consider first Problem I where $J(z)=1$. Then the most general solution of Eq.\ (\ref{exp1}) is
given by,
\begin{equation}
\psi_{\rm I}(z,s) = \frac{1}{(1-s)} + A_1(s)\, e^{-\sqrt{1-s}\, z} + B_1(s)\, e^{\sqrt{1-s}\, z},
\label{exp2}
\end{equation}
where $A_1(s)$ and $B_1(s)$ are two arbitrary $z$ independent constants. Since the solution
cannot diverge exponentially as $z\to \infty$, one gets $B_1(s)=0$. We need to still determine 
the constant $A_1(s)$. Here we use a method that is slightly different from that used in
Ref.\ \cite{CM}. We substitute the solution in Eq.\ (\ref{exp2}) into the integral equation
(\ref{wh1}). Performing the integration explicitly, one finds that the solution in Eq.\ (\ref{exp2})
satisfies the integral equation if and only if $A_1(s) = -[1-\sqrt{1-s}]/(1-s)$. Thus, the full solution is 
given by~\cite{CM}
\begin{equation}
\psi_{\rm I}(z,s) =\sum_{n=0}^{\infty} q_n(z) s^n= \frac{1}{(1-s)}-\frac{1-\sqrt{1-s}}{1-s}\, e^{-\sqrt{1-s}\, z}.
\label{exp3}
\end{equation}
One can then get the expected maximum from this explicit solution
by an integration and an expansion in powers of $s$~\cite{CM}
\begin{equation}
E[M_n]= \frac{2}{\sqrt{\pi}}\, \frac{\Gamma(n+3/2)}{\Gamma(n+1)}-1\simeq 2\sqrt{\frac{n}{\pi}}-1\quad\quad {\rm 
as}\,\,\, n\to\infty,
\label{exp4}
\end{equation}
which is of the same general form as in Eq.\ (\ref{cm1}) with $\sigma=\sqrt{2}$ and $c=1$. In addition,
it is also instructive to derive the solution $q_n(z)$ for large $n$ by analysing its generating function in Eq.\ 
(\ref{exp3}) in the vicinity of $s=1$. 
Taking the limits $s\to 1$ and $z\to \infty$ but keeping $z\,\sqrt{1-s}$ fixed, one can replace 
the
the generating function by a Laplace transform, and inverting the Laplace transform one gets
the scaling solution valid for large $n$
\begin{equation}
q_n(z) \simeq {\rm erf}\left(\frac{z}{\sqrt{4n}}\right) + \frac{1}{\sqrt{\pi n}}\, e^{-z^2/{4n}},
\label{exp5}
\end{equation} 
where ${\rm erf}(z)= 2\, {\pi}^{-1/2}\, \int_0^{z} e^{-u^2}\, du$ is the error function.
Note that the first term on the rhs of Eq.\ (\ref{exp5}) corresponds to the continuum solution
of the diffusion equation for a particle starting at $z>0$ and staying above an absorbing boundary 
at $0$, which is also the same as the cumulative probability that the maximum of a 
continuous-time Brownian motion
stays below $z$ up to time $t$, provided one makes the standard correspondence 
$\sigma^2 n = 2 D t$  
between the 
discrete step number $n$ and the continuous time $t$, $D$ being the diffusion constant for the
Brownian motion. The second term on the rhs of Eq.\ (\ref{exp5}) corresponds to the leading
correction due to discrete jumps and indeed is responsible for the constant $c$ in Eq.\ (\ref{cm1}).
This can be seen by substituting the scaling solution in Eq.\ (\ref{exp5}) in 
the exact relation, $E[M_n]= \int_0^{\infty} (1-q_n(z))dz$. Upon integrating, one recovers
the large-$n$ asymptotic solution in Eq.\ (\ref{exp4}) and one sees explicitly
that indeed the constant $c=1$ in Eq.\ (\ref{exp4}) arises from the integration of the second term
in Eq.\ (\ref{exp5}).

We now turn to Problem II where $J(z)= R+z$. Proceeding exactly as in the first case one
finds that the explicit solution of the differential equation (\ref{exp1}) is given by
\begin{equation}
\psi_{\rm II}(z,s) = \frac{R+z}{1-s} + A_2(s)\, e^{-\sqrt{1-s}\, z},
\label{fex1}
\end{equation}
where we have used the boundary condition that the solution cannot diverge exponentially
as $z\to \infty$. Substituting this solution in the integral equation (\ref{wh1}) fixes the
constant $A_2(s)$ and we get the full solution,
\begin{equation}
\psi_{\rm II}(z,s) = \sum_{n=0}^{\infty} Q_n(z) s^n = \frac{R+z}{1-s} + (1-R)\, 
\frac{1-\sqrt{1-s}}{(1-s)}\, 
e^{-\sqrt{1-s}\,z}.
\label{fex2}
\end{equation}
The behavior of $Q_n(z)$  for large $n$ can be derived by analysing the generating function near $s=1$.
In this case, one finds that for large $n$,
\begin{equation}
Q_n(z) \simeq (z+1) + (R-1)\, {\rm erf}\left(\frac{z}{\sqrt{4n}}\right)+ \frac{R-1}{\sqrt{\pi n}}\, e^{-z^2/{4n}}.
\label{fex3}
\end{equation}
Comparison with the asymptotic solution of Problem I in Eq.\ (\ref{exp5}) shows that in Problem II,
the solution approaches an $n$-independent stationary solution as $n\to \infty$
\begin{equation}
Q_{\infty} (z) = z + 1 \quad\quad \hbox{ for all  }  z\ge 0.
\label{fex4} 
\end{equation}
This solution is also independent of $R$; all terms containing $R$ in Eq.\ (\ref{fex3})
disappear in the long-time limit.
The corrections to this stationary solution for large $n$ have the scaling forms similar to Problem I.

In fact, we will show in the next section that quite generically, i.e. for arbitrary 
continuous and symmetric kernel $f(z)$, the solution $Q_n(z)$ for Problem II always approaches a 
stationary solution
$Q_{\infty}(z)$ which is, generically, a nontrivial function of $z$. However, for large $z$,
we will show that this stationary solution has a rather simple asymptotic,
\begin{equation}
Q_{\infty} (z) \simeq z + c' \quad\quad \hbox{ as }  z\to \infty,
\label{gex1}
\end{equation}
where the constant $c'$ will be shown to be exactly equal to $c$ in Eq.\ (\ref{c1}).
In particular, for the uniform distribution $f(z)$ given in Eq.\ (\ref{uni}) where
one can relate back to the original $3$-d flux problem, we will show that indeed this
same constant $c'=c$ appears as the  
extrapolation length 
in the expression for flux in Eq.\ (\ref{phi2}).
In the particular example of the exponential kernel $f(z)= \exp[-|z|]/2$, we see explicitly 
that indeed $c'=c=1$ by inspecting Eqs.\ (\ref{fex4}) and (\ref{exp4}).
This thus proves a special case of the general result $c'=c$ valid for
arbitrary continuous and symmetric kernel $f(z)$.
Note also that for this special case of the exponential kernel, the stationary
solution $Q_{\infty}(z)$ in Eq.\ (\ref{fex4}) actually retains its asymptotic form
in Eq.\ (\ref{gex1}) all the way down to $z=0$. This property, however, is rather special
to the exponential kernel. For a generic continuous and symmetric kernel, $Q_{\infty}(z)$ has a 
nontrivial
form for small $z$ as will be shown in the next section.

Another quantity of interest, as we will see later in a more general context, is the transient behavior
of $Q_n(0)$ for Problem II. It follows by subsituting $z=0$ in Eq.\ (\ref{fex2})
\begin{equation}
\sum_{n=0}^{\infty} Q_n(0) s^n = \frac{1}{1-s} + \frac{R-1}{\sqrt{1-s}}.
\label{fex5}
\end{equation}
Expanding the rhs of Eq.\ (\ref{fex5}) in powers of $s$ one gets
\begin{equation}
Q_n(0)= 1 + (R-1)\, {{2n}\choose n}\frac{1}{2^{2n}} \simeq 1 + \frac{(R-1)}{\sqrt{\pi n}}\quad\quad {\rm as}\,\,\, n\to \infty.
\label{fex6}
\end{equation}
We will see later that this transient behavior for the exponential kernel confirms, as a special case, the
validity of a general result in Eq.\ (\ref{trans6}) proved for arbitrary continuous and symmetric 
kernels.

\section{General Solution to the Wiener-Hopf problem}

In this section, we present an explicit solution to the integral equation (\ref{wh1})
for the two different inhomogeneous terms in Eqs.\ (\ref{rwp}) and (\ref{flp}). Our result
is valid for any arbitrary continuous and symmetric kernel $f(z)$ that is normalized, 
$\int_{-\infty}^{\infty} f(z)dz=1$
and with a finite second moment $\sigma^2= \int_{-\infty}^{\infty} z^2 f(z) dz$.
Our method relies on a beautiful general formalism developed by Ivanov~\cite{Ivanov} to deal
with half-space problems in the context of photon scattering. Let us first summarize
this formalism. Consider the integral equation (\ref{wh1}) with an arbitrary source
term $J(z)$. There are three steps to obtain the solution.

\begin{enumerate}

\item The first step is to define a Green's function $G(z,z_1,s)$ that satisfies the same integral 
equation but 
with a delta function source term, i.e.
\begin{equation}
G(z,z_1, s) = s \int_0^{\infty} G(z', z_1, s) f(z-z') dz' + \delta(z-z_1).
\label{green1}
\end{equation}
It is then easy to see that the solution of the inhomogeneous equation (\ref{wh1}) is given by
\begin{equation}
\psi(z,s) = \int_0^{\infty} G(z,z_1,s) J(z_1) dz_1.
\label{green2}
\end{equation}

\item The next step is to determine the Green's function $G(z,z_1,s)$ that satisfies Eq.\ (\ref{green1}).
One first defines the double Laplace transform,
\begin{equation}
{\tilde G} (\lambda,\lambda_1,s) = \int_0^{\infty} dz\, e^{-\lambda z}\, \int_0^{\infty} dz_1\, 
e^{-\lambda_1 z_1}\,
G(z,z_1,s).
\label{dlt1}
\end{equation}
Ivanov showed that this double Laplace transform can be determined in closed form by solving
Eq.\ (\ref{green1}) and is given by a simple form~\cite{Ivanov}
\begin{equation}
{\tilde G} (\lambda,\lambda_1,s) = \frac{\phi(s,\lambda)\, \phi(s, \lambda_1)}{\lambda+\lambda_1},
\label{dlt2}
\end{equation}
where the function $\phi(s,\lambda)$ is the following Laplace transform
\begin{equation}
\phi(s,\lambda)= \int_0^{\infty} dz\, e^{-\lambda z}\, G(z,0,s)
\label{dlt3}
\end{equation}

\item The third step is to obtain an explicit expression~\cite{Ivanov} for the function 
$\phi(s,\lambda)$
\begin{equation}
\phi(s,\lambda)= \exp\left[-\frac{\lambda}{\pi}\, \int_0^{\infty} \frac{\ln \left(1-s {\hat 
f}(k)\right)}{\lambda^2+k^2}\, dk\right],
\label{phisr}
\end{equation}
where ${\hat f}(k) = \int_{-\infty}^{\infty} f(x)\, e^{i\,k\,x}\, dx$ is the Fourier transform
of the kernel $f(x)$. 

\end{enumerate} 

Substituting the explicit expression for $\phi(s,\lambda)$ from Eq.\ (\ref{phisr}) into Eq.\ 
(\ref{dlt2}),
one has an explicit expression for the double Laplace transform $G(\lambda, \lambda_1, s)$. By 
inverting
this double transform, one can obtain the Green's function $G(z,z_1,s)$, at least in principle.
Subsequently, by performing the integral in Eq.\ (\ref{green2}) one obtains the required solution
$\psi(z,s)$. In practice, however, these last two steps are difficult to carry out explicitly
in general. However, for the two special source terms in Eqs.\ (\ref{rwp}) and (\ref{flp}), we show below
that one can make progress.

\subsection{General solution for Problem I}

Consider first Problem I where $J(z)=1$. Then, Eq.\ (\ref{green2}) gives
\begin{equation}
\psi_{\rm I}(z,s) =\int_0^{\infty} G(z,z_1,s) dz_1.
\label{p1g1}
\end{equation}
Let us define the Laplace transform
\begin{equation}
{\tilde \psi_{\rm I}}(\lambda,s)= \int_0^{\infty} \psi_{\rm I}(z,s)\, e^{-\lambda z}\, dz.
\label{p1lt}
\end{equation}
Taking the Laplace transform with respect to $z$ in Eq.\ (\ref{p1g1}) we get
\begin{equation}
{\tilde \psi_{\rm I}}(\lambda,s)= \int_0^{\infty}dz\, e^{-\lambda z}\, \int_0^{\infty} G(z,z_1,s)\, 
dz_1={\tilde G} (\lambda, 0, s).
\label{p1g2}
\end{equation}
Eqs.\ (\ref{dlt2}) and (\ref{phisr}) then give
\begin{equation}
{\tilde \psi_{\rm I}}(\lambda,s) =\frac{1}{\lambda}\, {\phi(s,\lambda)\, \phi(s,0)}
\label{p1g3}
\end{equation}
where $\phi(s,\lambda)$ is given in Eq.\ (\ref{phisr}). Let us first evaluate
$\phi(s,0)$. Note that one cannot na\"\i vely put $\lambda=0$ in the expression in Eq.\ (\ref{phisr})
since the integral multiplying $\lambda$ inside the exponential in Eq.\ (\ref{phisr}) diverges
as $\lambda\to 0$. Hence one needs to extract the value of $\phi(s,0)$ carefully. To achieve this,
an alternate expression for $\phi(s,\lambda)$ that was obtained in Ref.\ \cite{CM} turns out
to be useful. It was shown in Ref.\ \cite{CM} that $\phi(s,\lambda)$ in Eq.\ (\ref{phisr}) 
can also be written as
\begin{equation}
\phi(s,\lambda)= \frac{1}{\left[\sqrt{1-s}+\sigma\lambda \sqrt{s/2}\right]}\, 
\exp\left[-\frac{\lambda}{\pi}
\int_0^{\infty}
\frac{dk}{\lambda^2+k^2}\, \ln\left(\frac{1-s {\hat f}(k)}{1-s + s\sigma^2k^2/2}\right)\right].
\label{ana2}
\end{equation}
This representation is useful to derive the properties of $\phi(s,\lambda)$ near $\lambda=0$. On the
other hand, the representation in Eq.\ (\ref{phisr}) is useful to extract the asymptotic 
behavior of $\phi(s,\lambda)$ for large $\lambda$. Taking $\lambda\to 0$ limit in Eq.\ (\ref{ana2})
one gets
\begin{equation}
\phi(s,0)= \frac{1}{\sqrt{1-s}},
\label{p1g4}
\end{equation}
which, when substituted in Eq.\ (\ref{p1g3}) gives
\begin{equation}
{\tilde \psi_{\rm I}}(\lambda,s)=\sum_{n=0}^{\infty} s^n \int_0^{\infty} q_n(z)\, e^{-\lambda z}\, 
dz=
\frac{1}{\lambda\, \sqrt{1-s}}\, \phi(s,\lambda)
\label{p1g5}
\end{equation}
where $\phi(s,\lambda)$ is defined in Eq. (\ref{phisr}) and has also an alternative
expression as in Eq. (\ref{ana2}). This result in Eq. (\ref{p1g5}) goes
by the name of the Pollaczek-Spitzer formula which was originally
derived using completely 
different methods~\cite{Pollaczek,Spitzer1}. This result was subsequently
utilized in Ref.\ \cite{CM} to extract the constant $c$ in Eq.\ (\ref{cm1}) appearing as a subleading
term for large $n$ in the expected maximum $E[M_n]$ of a random walker.

\subsection{ General Solution for Problem II}

We now turn to Problem II where $J(z)=R+z$. We get from Eq.\ (\ref{green2})
\begin{equation}
\psi_{\rm II}(z,s) =\int_0^{\infty}(R+z_1) G(z,z_1,s) dz_1. 
\label{p2g1}
\end{equation}
Taking the Laplace transform, ${\tilde \psi_{\rm II}}(\lambda,s)= \int_0^{\infty} \psi_{\rm 
II}(z,s)\, 
e^{-\lambda z}\, dz$
gives
\begin{equation}
{\tilde \psi_{\rm II}}(\lambda,s)= \int_0^{\infty}dz\, e^{-\lambda z}\, \int_0^{\infty} (R+z_1)\, 
G(z,z_1,s)\, 
dz_1=R\, {\tilde G} (\lambda,0, s) - \frac{\partial {\tilde G}(\lambda,\lambda_1,s)}{\partial 
\lambda_1}|_{\lambda_1=0}.
\label{p2g2}
\end{equation}
Eqs.\ (\ref{dlt2}) and (\ref{phisr}) then give
\begin{equation}
{\tilde \psi_{\rm II}}(\lambda,s)= \frac{1}{\lambda}\,\left[ 
\left(R+\frac{1}{\lambda}\right)\,\phi(s,0)-
\frac{\partial {\tilde \phi}(s,\lambda_1)}{\partial \lambda_1}|_{\lambda_1=0}\right]\,\phi(s,\lambda)
\label{p2g3}
\end{equation}
where $\phi(s,\lambda)$ is given in Eq.\ (\ref{phisr}) or alternately in Eq.\ (\ref{ana2}). Using the
representation in Eq.\ (\ref{ana2}) one gets
\begin{equation}
\frac{\partial {\tilde \phi}(s,\lambda_1)}{\partial \lambda_1}|_{\lambda_1=0}= 
-\frac{\sigma}{(1-s)}\,\sqrt{\frac{s}{2}}
-\frac{1}{\pi \sqrt{1-s}}\int_0^{\infty} \frac{dk}{k^2}\, \ln\left(\frac{1-s {\hat f}(k)}{1-s + 
s\sigma^2k^2/2}\right).
\label{partial1}
\end{equation}
Substituting this result in Eq.\ (\ref{p2g3}) and using $\psi(s,0)=1/\sqrt{1-s}$ from Eq.\ (\ref{p1g4})
gives
\begin{equation}
{\tilde \psi_{\rm II}}(\lambda,s)= \frac{1}{\lambda}\,\left[ 
\frac{1}{\sqrt{1-s}}\left(R+\frac{1}{\lambda}\right) 
+ 
\frac{\sigma}{(1-s)}\,\sqrt{\frac{s}{2}} + \frac{1}{\pi \sqrt{1-s}}\int_0^{\infty} \frac{dk}{k^2}\, 
\ln\left(\frac{1-s {\hat f}(k)}{1-s +
s\sigma^2k^2/2}\right)\right]\, \phi(s,\lambda).
\label{p2g4}
\end{equation}

\subsection{Analysis of the Results}

Thus the Laplace transform of the solution to the integral equation (\ref{wh1}) for both
problems respectively with $J(z)=1$ (Problem I) and $J(z)=R+z$ (Problem II) can be written 
as a product of two functions 
\begin{equation}
{\tilde \psi}(\lambda,s)= W(s, \lambda)\, \phi(s,\lambda) 
\label{ana3}
\end{equation}
where $\phi(s,\lambda)$ is given in Eq.\ (\ref{phisr}) or in Eq.\ (\ref{ana2}) and is independent
of the source term. The function $W(s,\lambda)$, however, depends explicitly on the source term,
i.e. on the initial conditions of the original recursion relations in Eqs.\ (\ref{hop1}) and
(\ref{fop4}) and has different expressions for the two problems. While for
Problem I it is rather simple 
\begin{equation}
W_I(s,\lambda)=\frac{1}{\lambda \sqrt{1-s}},
\label{p1w}
\end{equation}
for Problem II it has a more complicated expression
\begin{equation}
W_{II}(s,\lambda) = 
\frac{1}{\lambda}\,\left[ \frac{1}{\sqrt{1-s}}\left(R+\frac{1}{\lambda}\right) +
\frac{\sigma}{(1-s)}\,\sqrt{\frac{s}{2}} + \frac{1}{\pi \sqrt{1-s}}\int_0^{\infty} \frac{dk}{k^2}\,
\ln\left(\frac{1-s {\hat f}(k)}{1-s +
s\sigma^2k^2/2}\right)\right] 
\label{p2w}
\end{equation}

Since the function $\phi(s,\lambda)$ appears in both problems, it is useful to first list
its asymptotic properties for small and large $\lambda$. For small $\lambda$, it is 
convenient to use the representation in Eq.\ (\ref{ana2}). On the other hand, for large $\lambda$, 
the
representation in Eq.\ (\ref{phisr}) turns out to be more convenient. We find
\begin{eqnarray}
\phi(s,\lambda) & \simeq & \frac{1}{\sqrt{1-s}} - \alpha(s)\lambda + O(\lambda^2)\quad\quad {\rm 
as}\,\, 
\lambda\to 0 
\label{sr} \\
& \simeq & 1- \frac{\beta(s)}{\lambda} + O(\lambda^{-2}) \quad\quad {\rm as}\,\, \lambda\to \infty
\label{lr}
\end{eqnarray}
where the two functions $\alpha(s)$ and $\beta(s)$ are given by
\begin{eqnarray}
\alpha(s)& =& \frac{\sigma}{(1-s)}\,\sqrt{\frac{s}{2}}+\frac{1}{\pi \sqrt{1-s}}\int_0^{\infty} 
\frac{dk}{k^2}\, 
\ln\left(\frac{1-s {\hat f}(k)}{1-s +
s\sigma^2k^2/2}\right) \label{alpha}\\
\beta(s) &=& \frac{1}{\pi} \int_0^{\infty} dk\, \ln\left[1-s {\hat f}(k)\right].
\label{beta}
\end{eqnarray}

Now we are ready to obtain the constants $c$ in Problem I and $c'$ in Problem II and
show that indeed they are same, as they both emerge from the properties of the function
$\phi(s,\lambda)$ that is common to both the problems. We consider
the two problems separately.
\vspace{0.4cm}

\noindent{\bf Expected maximum in Problem I:} Consider first Problem I. Consider 
the
Laplace transform of the distribution of the maximum, $E[e^{-\lambda M_n}]\equiv \int_0^{\infty} 
e^{-\lambda\,z} 
q_n'(z)\, dz= \lambda\int_0^{\infty} e^{-\lambda\, z} q_n(z)\,dz$. It follows from Eq.\ (\ref{p1g5})
\begin{equation}
\sum_{n=0}^{\infty} s^n E[e^{-\lambda M_n}]= \frac{1}{\sqrt{1-s}}\, \phi(s,\lambda).
\label{ps1}
\end{equation}
We expand both sides in $\lambda$ for small $\lambda$ and use Eq.\ (\ref{sr}) for the expansion of 
the 
rhs
of Eq.\ (\ref{ps1}). Comparing the term linear in $\lambda$ one gets
\begin{equation}
\sum_{n=0}^{\infty} s^n E[M_n]= \frac{\alpha(s)}{\sqrt{1-s}},
\label{ps2}
\end{equation}
where $\alpha(s)$ is given in Eq.\ (\ref{alpha}). To extract the large $n$ behavior of $E[M_n]$, we
need to analyse the generating function in Eq.\ (\ref{ps2}) near $s\to 1$. Expanding the rhs of
Eq.\ (\ref{ps2}) near $s=1$ one gets
\begin{equation}
\sum_{n=0}^{\infty} s^n E[M_n] \simeq \frac{\sigma}{\sqrt{2}\, (1-s)^{3/2}} + \frac{1}{\pi 
(1-s)}\int_0^{\infty} \frac{dk}{k^2} \ln\left[\frac{(1-{\hat f}(k)}{\sigma^2k^2/2}\right] +O((1-s)^{-1/2}).
\label{ps3}
\end{equation}
Inverting one gets the large $n$ result~\cite{CM} 
\begin{equation}
E[M_n]= \sigma \sqrt{\frac{2n}{\pi}} - c + O(n^{-1/2}),
\label{ps4}
\end{equation}
where the constant $c$ is given by the integral in Eq.\ (\ref{c1}).

\vspace{0.4cm}

\noindent{\bf Steady-state density profile in Problem II:} Turning now to Problem II, we first show 
that for large $n$, $Q_n(z)$ approaches a 
stationary solution $Q_{\infty}(z)$. From the definition, we have
\begin{equation}
\sum_{n=0}^{\infty} s^n \int_0^{\infty} Q_n(z)\, e^{-\lambda \, z}\, dz = {\tilde \psi_{\rm 
II}}(\lambda, s),
\label{zs1}
\end{equation}
where ${\tilde \psi_{\rm II}}(\lambda, s)$ is given in Eq.\ (\ref{p2g4}). Thus, if $Q_n(z)\to 
Q_{\infty}(z)$
as $n\to \infty$, then the lhs of Eq.\ (\ref{zs1}) will behave as
\begin{equation}
\sum_{n=0}^{\infty} s^n \int_0^{\infty} Q_n(z)\, e^{-\lambda \, z}\, dz \simeq \frac{1}{1-s}\, 
\int_0^{\infty} 
Q_{\infty} (z) e^{-\lambda\, z}\, dz \quad\quad {\rm as } \,\,\, s\to 1.
\label{zs2}
\end{equation}
We now expand the rhs of Eq.\ (\ref{zs1}) near $s=1$ using the explicit expression of 
${\tilde \psi_{\rm II}}(\lambda, s)$ in Eq.\ (\ref{p2g4}). We find from Eq.\ (\ref{p2g4}) that as 
$s\to 1$, 
the leading 
order term behaves as
\begin{equation}
{\tilde \psi_{\rm II}}(\lambda, s) \simeq \frac{\sigma\, \phi(1,\lambda)}{\sqrt{2}\lambda\, (1-s)}.
\label{zs3}
\end{equation}
Comparing Eqs.\ (\ref{zs2}) and (\ref{zs3}) gives the exact Laplace transform of the stationary solution
\begin{equation}
\int_0^{\infty} Q_{\infty} (z) e^{-\lambda\, z}\, dz = \frac{\sigma\, \phi(1,\lambda)}{\lambda\, 
\sqrt{2}},
\label{stat0}
\end{equation}
where $\phi(1,\lambda)$ can be obtained by putting $s=1$ either in Eq.\ (\ref{phisr}) or alternately
in Eq.\ (\ref{ana2}). Both expressions are equivalent and one gets
\begin{eqnarray}
\int_0^{\infty} Q_{\infty} (z) e^{-\lambda\, z}\, dz 
&=& \frac{\sigma}{\lambda\sqrt{2}}\,
\exp\left[-\frac{\lambda}{\pi}\,\int_0^{\infty} \frac{dk}{\lambda^2+k^2}\, \ln\left(1-{\hat 
f}(k)\right)\right]
\label{stat1}\\
&=& \frac{1}{\lambda^2} \exp\left[-\frac{\lambda}{\pi}\,\int_0^{\infty} \frac{dk}{\lambda^2+k^2}\,
\ln\left(\frac{1- {\hat f}(k)}{\sigma^2k^2/2}\right)\right] .
\label{stat2}
\end{eqnarray}
As a check, one can verify that for the exponential kernel $f(x)=\exp[-|x|]/2$ so that
${\hat f}(k)= 1/(1+k^2)$ and $\sigma=\sqrt{2}$, Eq.\ (\ref{stat2}) gives
\begin{equation}
\int_0^{\infty}
Q_{\infty} (z) e^{-\lambda\, z}\, dz = \frac{1}{\lambda^2} +\frac{1}{\lambda}
\label{stat2'}
\end{equation}
which, when inverted, gives $Q_{\infty}(z) =z+1$, in agreement with Eq.\ (\ref{fex4}).

For a general kernel $f(z)$, it is difficult to invert the Laplace transform
in Eqs.\ (\ref{stat1}) or (\ref{stat2}). However, one can easily extract the asymptotic behavior for large
and small $z$. Consider first the large $z$ behavior. In this case, the expression
in Eq.\ (\ref{stat2}) is more convenient. Expanding the rhs of Eq.\ (\ref{stat2}) for small $\lambda$ 
we get
\begin{equation}
\int_0^{\infty} Q_{\infty} (z) e^{-\lambda\, z}\, dz \simeq \frac{1}{\lambda^2}-\frac{1}{\lambda\, 
\pi}\, 
\int_0^{\infty}\frac{dk}{k^2}\,\ln \left[\frac{1-{\hat f}(k)}{\sigma^2k^2/2}\right].
\label{stat3}
\end{equation}
Inverting the Laplace transform, one gets 
\begin{equation}
Q_{\infty}(z) \simeq z + c' \quad\quad {\rm as}\,\,\, z\to \infty
\label{stat4}
\end{equation}
where 
\begin{equation}
c' = - \frac{1}{\pi}\, \int_0^{\infty} \frac{dk}{k^2}\, \ln \left[\frac{1-{\hat
f}(k)}{\sigma^2k^2/2}\right]=c
\label{stat5}
\end{equation} 
thus proving one of the main results of this paper. In particular, for the uniform kernel in Eq.\ 
(\ref{uni}) such that ${\hat f}(k)=\sin (k)/k$, we get
\begin{equation}
c'=c = 0.29795219028\dots
\label{stat5c}
\end{equation}
The asymptotic exact result $Q_{\infty} (z) \to z+ c'$ as $z\to \infty$ 
with $c'= 0.29795219028..$, when substituted in Eq. (\ref{denprof3}) in the
steady-state $n\to \infty$ limit, provides the
exact steady-state density profile in the
original $3$-d flux problem at distance $r>>R$,
\begin{equation}
\rho_{\infty}(r>>R)= \frac{\rho_0}{r}Q_{\infty}(r-R)=\rho_0\left(1-\frac{R-c'}{r}\right),
\label{denprof4}
\end{equation}
in perfect agreement with 
the numerically observed~\cite{Ziff1} behavior in Eq. (\ref{denprof2}) (note that
we have set $l=1$ in Eq. (\ref{denprof2})).

Similarly, one can 
work out the small $z$ behavior of $Q_{\infty}(z)$ by analysing the
large $\lambda$ behavior of the Laplace transform. In this case, the expression in Eq.\ (\ref{stat1})
is more convenient. Expanding Eq.\ (\ref{stat1}) for large $\lambda$ one finds
\begin{equation}
\int_0^{\infty} Q_{\infty} (z) e^{-\lambda\, z}\, dz =
\frac{\sigma}{\sqrt{2}}\,\left[\frac{1}{\lambda} -\frac{1}{\pi \lambda^2} \int_0^{\infty} dk\, 
\ln\left(1- {\hat 
f}(k)\right)+ O(\lambda^{-3})\right]
\label{stat6}
\end{equation}
Inverting the Laplace transform, we get 
\begin{equation}
Q_{\infty}(z) \simeq \frac{\sigma}{\sqrt{2}} + b z \quad\quad {\rm as}\,\,\, z\to 0,
\label{stat7}
\end{equation}
where $b$ is a new constant given by
\begin{equation}
b = -\frac{\sigma}{\pi\sqrt{2}} \int_0^{\infty} dk\, \ln\left(1- {\hat f}(k)\right).
\label{bex1}
\end{equation}
For the exponential case, ${\hat f}(k)=1/(k^2+1)$ and $\sigma=\sqrt{2}$ we get from Eq.\ (\ref{bex1}) 
\begin{equation}
b = \frac{1}{\pi}\int_0^{\infty} dk\, \ln \left[\frac{(k^2+1)}{k^2}\right]=1.
\label{bex2}
\end{equation}
Thus from Eq.\ (\ref{stat7}), for the exponential kernel, we get
$Q_{\infty}(z)= z+1 $ as $z\to 0$, in agreement with Eq.\ (\ref{fex4}).
For the uniform case, ${\hat f}(k)=\sin(k)/k$ and $\sigma=1/\sqrt{3}$, we get
\begin{equation}
b = -\frac{1}{\pi \sqrt{6}}\int_0^{\infty} dk \, \ln\left[1-\frac{\sin (k)}{k}\right]=0.653857\dots
\label{bex3}
\end{equation}
Thus, for the uniform kernel, the small $z$ behavior of $Q_{\infty}(z)$ from Eq.\ (\ref{stat7})
reads as
\begin{equation}
Q_{\infty}(z) \simeq \frac{1}{\sqrt{6}} + (0.653857\dots)\, z \quad\quad {\rm as}\,\,\, z\to 0
\label{bex4}
\end{equation}
This is in perfect agreement with the numerical results obtained from the work of Ref.\ \cite{Ziff1}, 
$Q_{\infty}(z) \approx 0.408245 + 0.6538 z $ as $z\to 0$~\cite{note1}. Substituting the
result from Eq. (\ref{bex4}) in Eq. (\ref{denprof3}) in the limit $n\to \infty$, we find
the steady-state density profile in the original $3$-d flux problem near the surface
$r\to R$
\begin{equation}
\rho_{\infty}(r\to R)= \frac{\rho_0}{r} Q_{\infty}(r-R)\approx \frac{\rho_0}{R}\left(\frac{1}{\sqrt{6}} 
+ 0.653857\cdots (r-R)\right).
\label{densurf2}
\end{equation}
In particular, on the surface, the steady-state density approaches to a constant value
\begin{equation}
\rho_{\infty}(R)= \frac{1}{\sqrt{6}}\, \frac{\rho_0}{R},
\label{densurf3}
\end{equation}
thus proving the numerically observed~\cite{Ziff1} relation in Eq. (\ref{densurf1}) 
and identifying the constant $0.408245$ as $1/\sqrt{6}$ (note that we have 
set $l=1$ in Eq. (\ref{densurf1})).  

\vspace{0.4cm}

\noindent{\bf Transient behavior in Problem II:}
Another quantity that was investigated numerically in Ref.\ \cite{Ziff1} is the transient behavior
of $Q_n(0)$ for large $n$, where it was found that for the uniform kernel in Eq.\ (\ref{uni})
\begin{equation}
Q_n(0)\approx 0.408245 +\left(\frac{R}{\sqrt{\pi}}-0.17\right)\, n^{-1/2}
\label{trans1}
\end{equation}
Our exact solution reproduces this result also. To see this, we investigate our explicit solution in Eq.\ 
(\ref{p2g4}) for a general continuous and symmetric kernel. 
By taking $\lambda\to \infty$ limit in Eq.\ (\ref{zs1}) we see that the lhs behaves as
\begin{equation}
\sum_{n=0}^{\infty} s^n \int_0^{\infty} Q_n(z)\, e^{-\lambda \, z}\, dz \simeq 
\frac{1}{\lambda}\sum_{n=0}^{\infty} 
Q_n(0) s^n \quad\quad {\rm as} \,\,\, \lambda\to \infty.
\label{trans2}
\end{equation}
On the other hand, expanding the exact expression of ${\tilde \psi_{\rm II}}(\lambda, s)$ in Eq.\ 
(\ref{p2g4})
for large $\lambda$, we find that the rhs of Eq.\ (\ref{zs1}) behaves as
\begin{equation}
{\tilde \psi_{\rm II}}(\lambda, s)\simeq \frac{1}{\lambda}\left[\alpha(s)+ 
\frac{R}{\sqrt{1-s}}\right]\quad\quad {\rm as}\,\,\, 
\lambda\to \infty
\label{trans3}
\end{equation}
where $\alpha(s)$ is given in Eq.\ (\ref{alpha}). Equating Eqs.\ (\ref{trans2}) and (\ref{trans3})
and using the expression for $\alpha(s)$ from Eq. (\ref{alpha})
we get an exact expression for the generating function
\begin{equation} 
\sum_{n=0}^{\infty} Q_n(0) s^n = \frac{\sigma}{(1-s)}\,\sqrt{\frac{s}{2}}+\frac{1}{\pi 
\sqrt{1-s}}\int_0^{\infty}
\frac{dk}{k^2}\,
\ln\left(\frac{1-s {\hat f}(k)}{1-s +
s\sigma^2k^2/2}\right) + \frac{R}{\sqrt{1-s}}.
\label{trans4}
\end{equation}
Now, analysing the behavior near $s=1$ of the rhs of Eq.\ (\ref{trans4}), one can get the
leading asymptotic behavior of $Q_n(0)$ for large $n$. One gets the leading behavior near $s=1$,
\begin{equation}
\sum_{n=0}^{\infty} Q_n(0) s^n \simeq \frac{\sigma}{\sqrt{2}\, (1-s)} + \frac{(R-c)}{\sqrt{1-s}}  
\label{trans5}
\end{equation}
where $c$ is the same constant as in Eq.\ (\ref{c1}).
Inverting, one obtains an exact asymptotic result for large $n$
\begin{equation}
Q_n(0)\simeq \frac{\sigma}{\sqrt{2}} + \frac{(R-c)}{\sqrt{\pi}}\, n^{-1/2} 
\label{trans6}
\end{equation}
In particular, for the uniform kernel ${\hat f}(k)=\sin(k)/k$ with $\sigma=1/\sqrt{3}$
and $c=0.29795219028\ldots$ from Eq.\ (\ref{stat5c}) we get
\begin{equation}
Q_n(0)\simeq \frac{1}{\sqrt{6}}+ \frac{(R-0.29795219028\ldots)}{\sqrt{\pi}}\, n^{-1/2}= 0.40824829\ldots 
+ \left(\frac{R}{\sqrt{\pi}}-0.168101522\ldots\right) \, n^{-1/2},
\label{trans7}
\end{equation}
in excellent agreement with the numerical results~\cite{Ziff1} stated in Eq.\ (\ref{trans1}).

\section{Sparre Andersen Theorem for Problem I and its Analogue for Problem II}

The recursion relation in Eq.\ (\ref{hop1}) satisfied by the $q_n(z)$'s in Problem I has
an explicit solution given in Eq.\ (\ref{p1g5}). From this explicit solution, one can
easily extract $q_n(0)$, the probability that a random walker starting at the origin stays
positive (or negative) up to $n$ steps. Indeed, the integral
$\int_0^{\infty} q_n(z) e^{-\lambda z} dz \to q_n(0)/\lambda$ in the $\lambda\to \infty$ limit.
On the other hand, the rhs of Eq.\ (\ref{p1g5}) tends to $1/{\lambda \sqrt{1-s}}$ as $\lambda\to 
\infty$
since $\phi(s, \infty)=1$. Matching the lhs and the rhs gives,
\begin{equation}
\sum_{n=0}^{\infty} q_n(0) s^n = \frac{1}{\sqrt{1-s}}.
\label{sa1}
\end{equation}
Expanding the rhs of Eq.\ (\ref{sa1}) in powers of $s$ and identifying the coefficient
of $s^n$ on both sides, one gets for all $n$
\begin{equation}
q_n(0) = {{2n}\choose n}\frac{1}{2^{2n}}.
\label{sa2}
\end{equation}
The amazing fact is that the result in Eq.\ (\ref{sa2}) is {\em universal} for all $n$, i.e. it
does not depend on the density function $f(z)$ as long as $f(z)$ is continuous
and symmetric.
This, in fact, is the celebrated Sparre Andersen theorem which was originally proved    
using combinatorial methods~\cite{SA} and has since been reproduced by various other
methods~\cite{Spitzer1,FF}. In particular, one notes from Eq.\ (\ref{sa2}) that in the 
limit of large $n$,
\begin{equation}
\lim_{n\to \infty} \sqrt{n}\,\, q_n(0) = \frac{1}{\sqrt{\pi}}= {\rm a\,\, universal\,\, constant}
\label{sa3}
\end{equation}
 
A natural question is if there is an analogue of the universality \` a la Sparre Andersen theorem
for Problem II. The recursion relation for Problem II in Eq.\ 
(\ref{fop4}) is identical
to that of Problem I in Eq.\ (\ref{hop1}), except that the initial condition $Q_0(z)=R+z$ is
different from that in Eq.\ (\ref{hop1}). The question is whether $Q_n(0)$ still remains
universal with this different initial condition.
Indeed, the answer to this question is evident from our exact result in Eq.\ (\ref{trans4}). 
It is clear from Eq.\ (\ref{trans4}) that unlike in Problem I, $Q_n(0)$ in Problem II is not 
universal for all $n$ as it depends explicitly on the density function ${\hat f}(k)$. However,
one recovers universality (up to a constant scale factor $\sigma$ ) asymptotically, i.e. in the limit
of large $n$. Indeed, it follows from Eq.\ (\ref{trans6}) 
\begin{equation}
\lim_{n\to \infty} \frac{Q_n(0)}{\sigma} = \frac{1}{\sqrt{2}}= {\rm a\,\, universal\,\, constant}
\label{sa4}
\end{equation}
Thus $Q_n(0)$, scaled by $\sigma$, approaches a universal constant $1/\sqrt{2}$ as $n\to \infty$,
independently of the density function $f(z)$ as long as $f(z)$ is continuous, symmetric and has a 
finite
second moment $\sigma^2= \int_{-\infty}^{\infty} z^2\, f(z)\, dz $. The result in Eq.\ (\ref{sa4})
for Problem II can be thought of as an asymptotic analogue of the Sparre Andersen result in Eq.\ 
(\ref{sa3}) for Problem I.

\section{Conclusions}
We have shown that the two apparently different discrete-time random walk problems, the expected
maximum in one dimension and the three-dimensional flux to a trap, are intimately related in that they end 
up satisfying the
same recursion relation, but with different initial conditions.  We confirmed that the constant 
$c'$ in the flux problem is identical to the constant $c$ in the expected-maximum problem,
and thus provide an exact derivation for this constant which was previously found only numerically,
fourteen years ago.  We also find the surprisingly simple result that the steady-state
density in the flux problem reaches a constant at the boundary
of the sphere, 
$\rho_{\infty}(R) = \rho_0 l / (\sqrt{6} R)$ or equivalently
$Q_{\infty}(0)=1/\sqrt{6}$, and find explicitly both
the asymptotic time-dependent approach to that value, Eq.\ (\ref{trans7}), and the slope of the density
at the wall at $z=0$ in the steady state, Eq.\ (\ref{bex4}).  

For the flux problem, $c'$ represents the extrapolation length inside the boundary
where the steady-state solution far from
the sphere $\rho_{\infty}(r) = \rho_0(r - R + c' \ell)/r$ goes to zero.  That is, the solution far 
from the
sphere assumes the form of the solution to the diffusion equation, but with the effective
boundary somewhat inside the actual boundary.  Putting this expression into the formula
for the flux, $\Phi = 4 \pi r^2 D (d/dr) \rho_{\infty}(r)$, yields the leading term 
in Eq.\ (\ref{phi2}) (with $l = 1$).  We evaluate $\Phi$ for large $r$ where the solution is
valid, and the result is naturally independent of $r$ because it is in steady state.

In this paper, we have presented explicitly the steady-state density profile 
for the $3$-d discrete flux problem. As mentioned above, this solution
suffices to predict immediately the leading behavior of the flux
$\Phi(t)$ as $t\to \infty$ in Eq.\ (\ref{phi2}). 
To obtain the subleading time-dependent term given in Eq.\ (\ref{phi2}), it is necessary to 
study the large time asymptotic behavior of the integral in Eq.\ (\ref{f3}).
Indeed, using the asymptotic solution for $Q_n(z)$ derived
here, it is possible to calculate this subleading term explicitly; we have not presented
this calculation here and it will be published elsewhere~\cite{ZMC}.
It is worth remarking that in the
result found for the flux in Eq.\ (\ref{phi2}), the combination $R-c' l $ enters in
the time-dependent correction in the same form as in the steady-state term ---
that is, the extrapolation-length idea also figures in the asymptotically large-time
behavior of the system.

We have considered the three-dimensional flux problem here, in which 
particles undergo a Rayleigh flight of unit step length.  The resulting equation,
Eq.\ (\ref{fop4}), can also 
be interpreted as representing a one-dimensional flux problem where the particles
undergo a uniformly distributed jump, with an adsorbing boundary at $z = 0$.
For the one-dimensional interpretation, however, the initial condition is uniform
rather than linear as in (\ref{flp}).  (Call this Problem II').
Then Problems I and II' become identical, with $E[M_n]$ of the
former, Eq.\ (\ref{emn1}), corresponding to the total integrated
flux up to time $n$ of the latter.  For a uniform
jump distribution, Eq.\ (\ref{flaj1}) thus  gives the accumulated flux up to time $n$.
In Ref.\ \cite{Ziff1}, the quantity $a^{(2)}(n)$ of Eq.\ (10a)
 is equal to six times the flux at time $n$ for precisely this one-dimensional problem.
Summing the data for $a^{(2)}(n)/6$ up to time $n$
 (some of which data is presented in Table I of that paper),
 we indeed find that Eq.\ (\ref{flaj1}) (including $c$) is satisfied, with
 the next correction approximately equal to $0.0921 n^{-1/2}$.  In fact, the latter agrees
 with the prediction $(1/5)\sqrt{2/3\pi n} = 0.09213177\ldots n^{-1/2}$ given in Ref.\ \cite{CFFH}
and this prediction also agrees
 with the conjectured behavior of $a^{(2)}(n)$ given in Eq.\ (21c) of Ref.\ \cite{Ziff1}.
 Thus, we have verified an additional
 conjecture of Ref.\ \cite{Ziff1}.

On the mathematical side, we have discussed explicit solutions to the 
Wiener-Hopf type integral
equations for continuous and symmetric kernel with a finite second moment
$\sigma^2$. This is because both the physics problems that we were interested in
this paper have kernels that satisfy these properties. 
Mathematically it is interesting to ask how the solutions would change if 
the kernel
is asymmetric or for example, does not have a finite second moment. This later
problem with diverging second moment corresponds to discrete-time L\'evy flights 
and at least for the expected maximum
problem (Problem I), exact results using similar techniques have recently
been obtained~\cite{CM}. On the other hand, while general solutions to
the Wiener-Hopf type integral equations with asymmetric kernels can, in principle,
be obtained~\cite{Ivanov}, they are mostly not explicit. Finding explicit solutions 
with asymmetric kernels remains a hard and challenging problem.     

Finally, our solution of the recursion relation in Eq. (\ref{hop1}) with a constant initial
condition includes, as a special case, a simple derivation of the Sparre Andersen theorem
that states that $q_n(0)$, i.e. the probability that starting at $0$ a random walker's path stays above
(or below) $0$ up to $n$ steps, is independent of the jump density function as long as it 
is continuous and symmetric. However, we have proved that for the same recursion 
relation, but with a 
linear
initial condition as in Eq. (\ref{fop4}), the universality in $Q_n(0)$ holds only
asymptotically for large $n$ up to a constant factor, i.e. the ratio $Q_n(0)/\sigma$ (where
$\sigma$ is the standard deviation associated with the jump 
density function)
approaches to a universal constant $1/\sqrt{2}$, irrespective of the details of the
jump density function as long as it is continuous and symmetric. 
This paper provides a rigorous proof of this new theorem.

\vspace{0.5cm}

{\bf Acknowledgments:} RMZ acknowledges support from NSF grant DMS-0244419.  
%xxx


\begin{thebibliography}{999}


\bibitem{CS} E.G. Coffman and P.W. Shor, Packing in two dimensions: asymptotic average-case analysis of 
algorithms, Algorithmica, {\bf 9}: 253-277 (1993).

\bibitem{CFFH} E.G. Coffman, P. Flajolet, L. Flato, and M. Hofri, The maximum of random walk
and its application to rectangle packing, Probability in Engineering
and Informational Sciences, {\bf 12}: 373-386 (1998).

\bibitem{CM} A. Comtet and S.N. Majumdar, Precise asymptotics for a random walker's maximum,
J. Stat. Mech.: Th. and Exp. {\bf P06013}: 1-17 (2005).

\bibitem{Ziff1} R.M. Ziff, Flux to a trap, J. Stat. Phys. {\bf 65}: 1217-1233 (1991).

\bibitem{Smolu} M. von Smoluchowski, Drei Vorträge über Diffusion, Brownsche 
Bewegung und Koagulation von Kolloidteilchen, Phys. Z. {\bf 17}: 557-571, 585-599 (1916).

\bibitem{Chandra} S. Chandrasekhar, Stochastic problems in physics and astronomy, Rev. Mod. Phys. {\bf 
15}: 1-89 (1943).

\bibitem{Weiss} G.H. Weiss, Overview of theoretical-models for reaction rates, J. Stat. 
Phys. {\bf 42}: 3-36 (1986).


\bibitem{BT} M.A. Burschka and U.M. Titulaer, The kinetic boundary layer theory for the Fokker-Planck
equation with absorbing boundary, J. Stat. Phys. {\bf 25}: 569-582 (1981).

\bibitem{CZ} K.M. Case and P.W. Zwiefel, {\em Linear Transport Theory} (Addison-Wesley, Reading, 
Massachusetts, 1967).

\bibitem{Williams} M.M.R. Williams, {\em The Slowing Down and Thermalization of Neutrons} (North-Holland, 
Amsterdam, 1966).

\bibitem{CK} F.C. Collins and G.E. Kimball, Diffusion-controlled reaction rates, J. Colloid Sci., {\bf 
4}: 425-437 (1949).

\bibitem{Redner} S. Redner, {\em A Guide to First-passage Processes}
(Cambridge University Press, Cambridge 2001), p 170.

\bibitem{Ivanov} V.V. Ivanov, Resolvent method: exact solutions of half-space transport problems by 
elementary means, Astron. Astrophys. {\bf 286}: 328-337 (1994).

\bibitem{Pollaczek} F. Pollaczek, Fonctions caract\'eristiques de certaines
r\'epartitions d\'efinies au moyen de la notion d'ordre,  Comptes rendus
{\bf 234}: 2334-2336 (1952).

\bibitem{Spitzer1} F. Spitzer, The Wiener-Hopf equation whose kernel is a probability density,
Duke Math. J. {\bf 24}: 327-343 (1957).

\bibitem{note1} The numerical value of the slope at $z=0$ in this formula was determined 
from the data
collected in the work in Ref.\ \cite{Ziff1}, but not published in that paper. We deduced
the value given here  before we determined the
theoretical result given in Eq.\ (\ref{bex4}). 

\bibitem{SA} E. Sparre Andersen, On the fluctuations of sums of random variables II, Matematica 
Scandinavica, {\bf 2}: 195-223 (1954).

\bibitem{FF} U. Frisch and H. Frisch, Universality of escape from a half-space for 
symmetrical random walks,
in {\em L\'evy Flights and Related Topics in Physics: Proceedings of the International Workshop 
held at Nice, France, Lecture Notes in Physics ed. {\rm M.F. Shlesinger et al. }}, 262-268 
(Springer-Verlag, 1994). 

\bibitem{ZMC} R.M. Ziff, S.N. Majumdar and A. Comtet, to be published.

\end{thebibliography}
\end{document}